\documentclass[final,3p,times,twocolumn]{elsarticle}
 \biboptions{comma,sort&compress}
\usepackage{here}
 \usepackage{graphicx}
  \usepackage{epsfig}

\newcommand{\bc}{\begin{center}}
\newcommand{\ec}{\end{center}}
\newcommand{\be}{\begin{equation}}
\newcommand{\ee}{\end{equation}}
\newcommand{\bea}{\begin{eqnarray}}
\newcommand{\eea}{\end{eqnarray}}
\newcommand{\ba}{\begin{array}}
\newcommand{\ea}{\end{array}}
\newcommand{\lb}{\label}
\newcommand{\rf}{\ref}
\newcommand{\bfg}{\begin{figure}[htbp]}
\newcommand{\efg}{\end{figure}}
\newcommand{\pr}{Phys. Rev. }
\newcommand{\np}{Nucl. Phys. }

\newcommand{\prp}{Phys. Rep. }

\newcommand{\pl}{Phys. Lett. }

\newcommand{\nc}{Nuovo Cimento }

\journal{Nuc. Phys. (Proc. Suppl.)}

\begin{document}

\begin{frontmatter}



\title{Properties of the gauge invariant quark Green's function 
\protect \\
in two-dimensional QCD}

 \author{H. Sazdjian}
  \address{Institut de Physique Nucl\'eaire, CNRS/IN2P3,\\
Universit\'e Paris-Sud 11, F-91405 Orsay, France}
\ead{sazdjian@ipno.in2p3.fr}

\begin{abstract}
\noindent
Using an exact integro\-differential equation we study the properties
of the gauge invariant quark Green's function, defined with a path-ordered 
gluon field phase factor along a straight line, in two-dimensional QCD in
the large-$N_c$ limit. The Green's function is found to be infrared 
finite with singularities represented by an infinite number of threshold 
type branch points with a power equal to -3/2, starting at positive mass 
squared values . The solution is analytically determined.
\end{abstract}


\end{frontmatter}


Gauge invariant quark Green's functions are defined with the
aid of path-ordered gluon field phase factors \cite{m,nm}.
Skew-polygonal lines for the paths are of particular interest
since they can be represented as junctions of simpler straight
line segments. For such lines with $n$ sides and $n-1$ junction 
points $y_1$, $y_2$, $\ldots$, $y_{n-1}$ between the segments, we 
define the Green's function $S_{(n)}$ as
\[S_{(n)}(x,x';y_{n-1},\dots,y_1)=
-\frac{1}{N_c}\,\langle\overline \psi(x')\,U(x',y_{n-1})\]
\be \lb{e1}
\ \ \ \ \ \times U(y_{n-1},y_{n-2})\ldots U(y_1,x)\,\psi(x)\rangle,
\ee
where $U(x,y)$ is a path-ordered phase factor along a straight line 
segment joining $y$ to $x$. The simplest such Green's function 
corresponds to $n=1$, for which the points
$x$ and $x'$ are joined by a single straight line:
\[S_{(1)}(x,x')\equiv S(x,x')=-\frac{1}{N_c}\,
\langle\overline \psi(x')\,U(x',x)\,\psi(x)\rangle.\]
\be \lb{e2}
\ee
(We shall generally omit the index 1 from that function.)
\par
The theory is quantized in two steps. First, one integrates with 
respect to the quark fields. This produces in various terms the 
quark propagator in the presence of the gluon field. Then one 
integrates with respect to the gluon field through Wilson loops 
\cite{w,p,mm1,mm2,mgd,mk}. To achieve the latter operation,
we use for the quark propagator in external field a
representation which involves phase factors along straight lines
together with the full gauge invariant quark Green's function
\cite{s1,js}. This representation is a generalization of the one 
introduced by Eichten and Feinberg when calculating the 
relativistic effects starting from a nonrelativistic limit \cite{ef}. 
\par
The quark propagator in the external gluon field is expanded
around the following gauge covariant quantity:
\be \lb{e3}  
\Big[\widetilde S(x,x')\Big]_{\ b}^a\equiv 
S(x,x')\Big[U(x,x')\Big]_{\ b}^a.
\ee
It is possible to set up an integral equation realizing the 
previous expansion. Its systematic use leads to the derivation of
functional relations between the Green's functions $S_{(n)}$ 
(skew-polygonal line with $n$ segments) and $S$ (one segment).
\par
Using then the equations of motion relative to the Green's functions, 
one establishes the following equation for $S(x,x')$ \cite{s1}:
\bea \lb{e4}
& &(i\gamma.\partial_{(x)}-m)S(x,x')=i\delta^4(x-x')\nonumber \\
& &\ \ +i\gamma^{\mu}\Big\{K_{2\mu}(x',x,y_1)\,S_{(2)}(y_1,x';x)
\nonumber \\
& &\ \ +\sum_{n=3}^{\infty}K_{n\mu}(x',x,y_1,\ldots,y_{n-1})\nonumber \\
& &\ \ \ \ \ \ \ \ \ 
\times S_{(n)}(y_{n-1},x';x,y_1,\ldots,y_{n-2})\Big\},
\eea
where the kernel $K_n$ ($n=2,3,\ldots$) contains globally $n$ 
derivatives of Wilson loop averages with skew-polygonal contours 
and also the Green's function $S$ and its derivative. (Integrations 
on intermediate variables are implicit.) The Green's functions 
$S_{(n)}$ being themselves related to the simplest Green's function 
$S$ through series expansions resulting from functional relations, 
(\rf{e4}) is ultimately an integro\-diffe\-rential equation for $S$. 
One expects that the kernels with small numbers of derivatives will 
provide the leading contributions. Therefore, the first kernel $K_2$ 
in (\rf{e4}) would contain the driving term of the interaction.  
\par
Equation (\rf{e4}) shows that the Green's function $S$ should have
singularities in momentum space, generated by the free quark propagator
(the inverse of the Dirac operator in the left-hand side). On the 
other hand, such singularities cannot be generated by saturating the
Green's function with intermediate states made of hadrons, which are
color singlets; the intermediate states that saturate the Green's 
function are necessarily colored states. It is therefore necessary 
to assume that quarks and gluons, which are colored objects, continue 
forming a complete set of states for saturation schemes and contribute, 
as the building blocks of the theory, with positive energies. It is only 
the solution of eq. (\rf{e4}) which should provide the indication about 
the issue of their physical status in observable phenomena. The above 
hypothesis has an immediate consequence about the analytic properties 
of the Green's function: it satisfies a generalized form of the 
K\"all\'en--Lehmann representation \cite{s1,kl,l,wght,schwb,thv}.
\par
As a first attempt to solve eq. (\rf{e4}), we have considered the case
of two-dimensional QCD in the large-$N_c$ limit \cite{th1,th2,ccg}. 
That theory is expected to have similar properties as four-dimensional 
QCD concerning its confining aspect; furthermore, asymptotic freedom
is realized there rather trivially, since the theory is 
super\-renor\-ma\-lizable. In two dimensions, the logarithm of the 
Wilson-loop average of a simple contour is proportional to the area 
enclosed by the contour \cite{kkk,k,b}. Equation (\rf{e4}) considerably 
simplifies and it is only the lowest-order kernel $K_2$ that survives. 
Additional simplification arises from the fact that it involves a 
functional second-order derivatve of the Wilson-loop average, which 
in two dimensions reduces to a delta-function. Taking also into account 
translation invariance, eq. (\rf{e4}) becomes \cite{s2}:
\[(i\gamma.\partial-m)S(x)=i\delta^2(x)\]
\[\ \ \ \ -\sigma\gamma^{\mu}(g_{\mu\alpha}g_{\nu\beta}
-g_{\mu\beta}g_{\nu\alpha})x^{\nu}x^{\beta}\]
\[\ \ \ \ \ \ \ \times\Big[\,\int_0^1d\lambda\,\lambda^2\,S((1-\lambda)x)
\gamma^{\alpha}S(\lambda x)\]
\be \lb{e5}
\ \ \ \ \ \ \ \ \ \ +\int_1^{\infty}d\xi\,S((1-\xi)x)
\gamma^{\alpha}S(\xi x)\,\Big], 
\ee
where $\sigma$ is the string tension.
\par
This equation is solved by decomposing $S$ into Lorentz invariant
parts:
\be \lb{e6}
S(p)=\gamma.pF_1(p^2)+F_0(p^2),
\ee
or, in $x$-space:
\be \lb{e7}
S(x)=\frac{1}{2\pi}\Big(\frac{i\gamma.x}{r}\widetilde F_1(r)
+\widetilde F_0(r)\Big),\ \ \ \ \ r=\sqrt{-x^2}. 
\ee
\par
One obtains, with the introduction of the Lorentz invariant functions,
two coupled equations. Their resolution proceeds through several steps, 
mainly based on the analyticity properties resulting from the
spectral representation of the Green's function \cite{s2}. The solutions 
are obtained in explicit form for any value of the quark mass $m$.
\par
The covariant functions $F_1(p^2)$ and $F_0(p^2)$ are, for complex $p^2$:
\bea
\lb{e8}
& &F_1(p^2)=-i\frac{\pi}{2\sigma}\,\sum_{n=1}^{\infty}\,
b_n\,\frac{1}{(M_n^2-p^2)^{3/2}},\\
\lb{e9}
& &F_0(p^2)=i\frac{\pi}{2\sigma}\,\sum_{n=1}^{\infty}\,
(-1)^nb_n\,\frac{M_n}{(M_n^2-p^2)^{3/2}}.
\eea
The masses $M_n$ ($n=1,2,\ldots$) have positive values greater than
the quark mass $m$ and are labelled with increasing values with respect 
to $n$; their squares represent the locations of branch point 
singularities with power $-3/2$. The masses $M_n$ and the coefficients
$b_n$ satisfy an infinite set of coupled algebraic equations that are
solved numerically. Their asymptotic behaviors for large $n$, such that
$\sigma\pi n\gg m^2$, are:
\be \lb{e10}
M_n^2\simeq \sigma\pi n,\ \ \ \ \ \ \ 
b_n\simeq \frac{\sigma^2}{M_n}.
\ee
\par
In $x$-space, the solutions are:
\be \lb{e11}
\widetilde F_1(r)=\frac{\pi}{2\sigma}\,\sum_{n=1}^{\infty}\,
b_n\,e^{-M_nr},
\ee
\be \lb{e12}
\widetilde F_0(r)=\frac{\pi}{2\sigma}\,\sum_{n=1}^{\infty}\,
(-1)^{n+1}b_n\,e^{-M_nr}.
\ee
[$r=\sqrt{-x^2}$.] 
\par
We present in Fig. \rf{f1} the function $iF_0$ for spacelike $p$ and 
in Fig. \rf{f2} its real part for timelike $p$, for the case $m=0$. 
\par
In conclusion, the spectral functions of the quark Green's function 
are infrared finite and lie on the positive real axis of $p^2$. No 
singularities in the complex plane or on the negative real axis have 
been found. This means that quarks contribute like physical particles 
with positive energies. (In two dimensions there are no physical gluons.)
\par 
\bfg
\bc
\epsfig{file=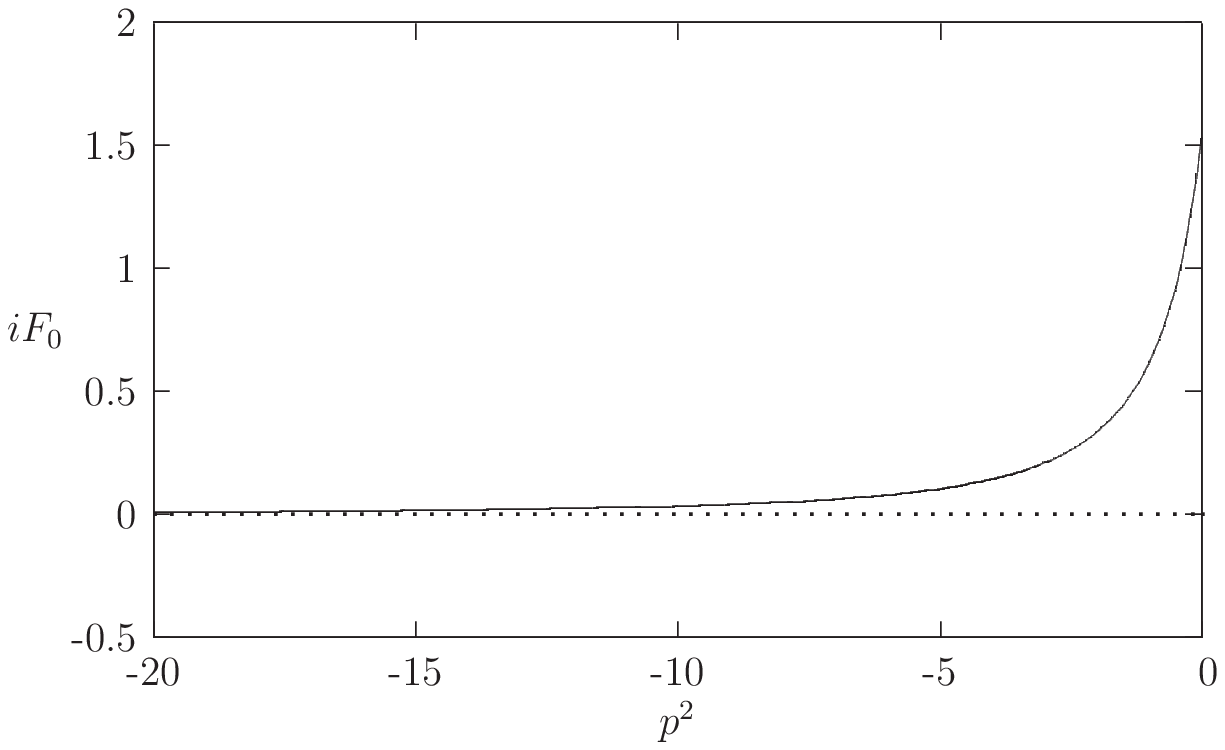,scale=0.48,bbllx=80,bblly=470,bburx=530,bbury=720}
\caption{The function $iF_0$ for spacelike $p$, in mass unit of 
$\sqrt{\sigma/\pi}$, for $m=0$.} 
\lb{f1}
\ec
\efg
\par
The singularities of the Green's function are represented by an
infinite number of threshold type singularities, characterized by 
a power of $-3/2$ and positive masses $M_n$ ($n=1,2,\ldots$). The
corresponding singularities are stronger than simple poles and this 
feature is an indication about the difficulty in the observability 
of quarks as asymptotic states.
\par
\bfg
\bc
\epsfig{file=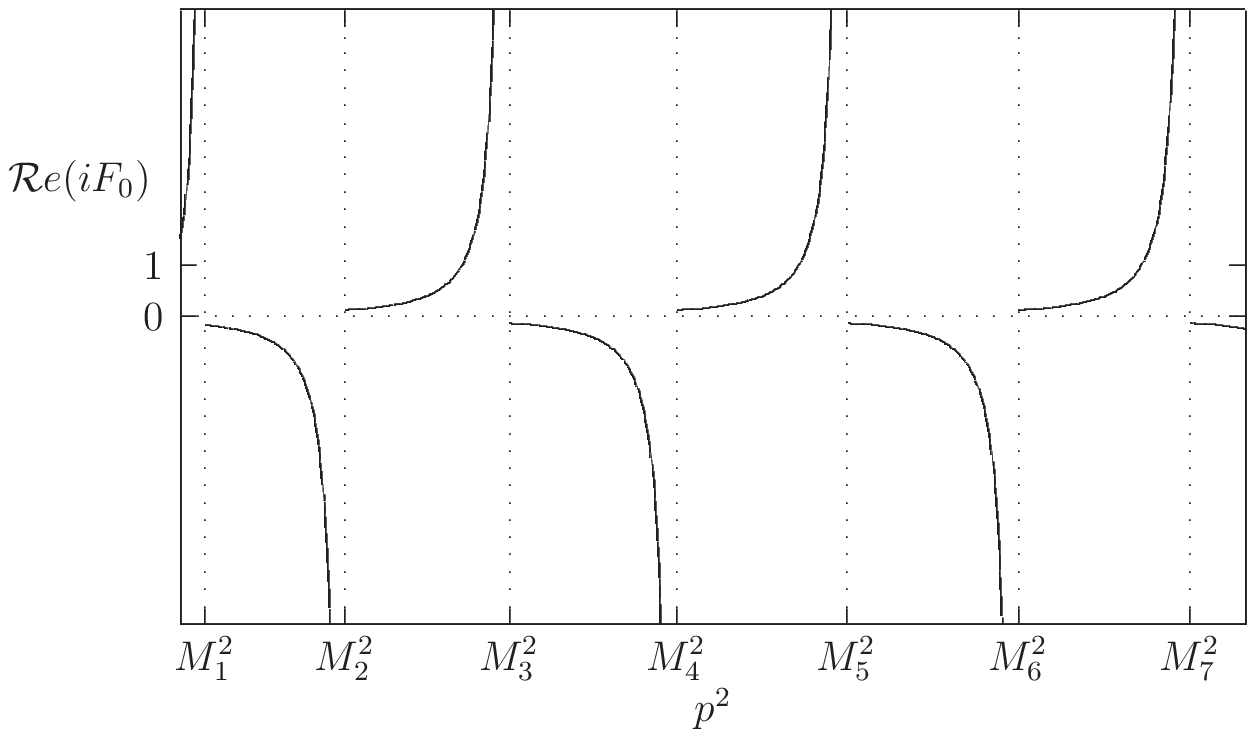,scale=0.48,bbllx=80,bblly=470,bburx=530,bbury=720}
\caption{The real part of the function $iF_0$ for timelike $p$, in mass 
unit of $\sqrt{\sigma/\pi}$, for $m=0$.} 
\lb{f2}
\ec
\efg
\par
The threshold masses $M_n$ represent dynamically generated masses and 
maintain the scalar part of the Green's function at a nonzero value
even when the quark mass is zero.
\par
\section*{Acknowledgements}
This work was supported in part by the EU network FLAVIANET, under 
Contract No. MRTN-CT-2006-035482, and by the European Community 
Research Infrastructure Integrating Activity ``Study of Strongly 
Interacting Matter'' (acronym HadronPhysics2, Grant Agreement 
No. 227431), under the Seventh Framework Programme of EU.


\end{document}